\documentclass[twocolumn,aps,prl,showpacs,preprintnumbers,amsmath,amssymb]{revtex4}
\usepackage{graphicx}% Include figure files
\usepackage{dcolumn}% Align table columns on decimal point
\usepackage{bm}% bold mathx
\usepackage{amssymb}%symbols of the figures
\usepackage{wasysym}%other symbols of the figures
\usepackage[rflt]{floatflt}
\usepackage{bibtopic}

\begin{document}

\preprint{APS}

\title {About the strength of correlation effects in the electronic structure of iron}

   \author{J. S\'anchez-Barriga$^{1}$, J. Fink$^{1,2}$, V. Boni $^3$, I. Di Marco$^{4,5}$, 
   J. Braun$^6$, J. Min\'ar$^6$,  A. Varykhalov$^1$, O. Rader$^1$, V. Bellini$^3$, 
   F. Manghi$^3$, H. Ebert$^6$, 
   M.I. Katsnelson$^5$, A. I. Lichtenstein$^7$, O. Eriksson$^4$, 
   W. Eberhardt$^1$, and H. A. D\"{u}rr$^1$\\}

\affiliation {$^1$Helmholtz-Zentrum Berlin f\"{u}r Materialien und Energie,
Elektronenspeicherring BESSY II, Albert-Einstein-Strasse 15, D-12489 Berlin, Germany\\
$^2$Leibniz-Institute for Solid State and Materials Research Dresden, P.O. Box 270116, D-01171 Dresden, Germany\\
$^3$Dipartimento di Fisica, Universit\`{a} di Modena, Via Campi 213/a, I-41100 Modena, Italy\\
$^4$Department of Physics and Material Science, Uppsala University, Box 530, SE-751 21 Uppsala, Sweden\\
$^5$Institute of Molecules and Materials, Radboud University of Nijmegen, Heijendaalsweg 135, 6525 AJ Nijmegen, The Netherlands\\
$^6$ Dep. Chemie und Biochemie, Physikalische Chemie, Universit\"{a}t M\"{u}nchen, Butenandtstr. 5-13, D-81377,M\"{u}nchen, Germany\\
$^7$Institute of Theoretical Physics, University of Hamburg, 20355 Hamburg, Germany}
\date{\today}

\begin{abstract}

The strength of electronic correlation effects in the spin-dependent electronic structure of 
ferromagnetic bcc Fe(110) has been investigated by means of spin and angle-resolved photoemission 
spectroscopy. The experimental results are 
compared to theoretical calculations within the three-body scattering approximation and within the 
dynamical mean-field theory, together with one-step model calculations of the photoemission process. 
This comparison indicates
that the present state of the art many-body calculations, although improving the description of correlation 
effects in Fe, give too small mass renormalizations and scattering rates thus demanding more refined 
many-body theories including non-local fluctuations.

\end{abstract}
\pacs{75.70.Rf, 79.60.Bm, 73.20.At, 71.15.Mb, 75.50.Bb}
\maketitle

Since more than half a century it is clear that the bandstructure together with exchange 
and correlation effects 
play an important role for the appearance of ferromagnetism in $3d$ transition metals and their 
alloys \cite{Herring-M-66}.
A first step toward an understanding of the electronic structure of these metals has been achieved by calculations
of the single-particle band dispersion [$E({\bf k})$] within the density functional theory (DFT) in the local 
spin-density approximation (LSDA) \cite{Moruzzi-C-78} which takes into account correlation 
effects only in a limited extent. It soon turned out that for the ferromagnetic 
$3d$ transition metals such as Fe, Co, and Ni, calculations 
beyond DFT-based theories have to be developed to take into account
many-body interaction, i.e., correlation effects, 
which normally are described by the energy and momentum
dependent complex self-energy function $\Sigma(E,\bf k)$. Here
the real part $\Re\Sigma$ is related to the mass enhancement while the imaginary part $\Im\Sigma$ 
describes the scattering
rate. One of the successful 
schemes for correlated electron systems
is the dynamical mean-field theory (DMFT). It replaces the problem of describing 
correlation effects in a  periodic lattice 
by a correlated impurity coupled to a  
self-consistent bath \cite{Georges-RMP-96}. An alternative approach is the three-body scattering (3BS) 
approximation which takes into account the scattering of a hole into an Auger-like excitation 
in the valence band,
formed by one hole plus an electron-hole excitation \cite{Calandra-PRB-94}. Such many-body 
calculations allowed the 
qualitative description of the  quenching of majority-channel quasiparticle 
excitations in Co \cite{Monastra-PRL-02} 
or the narrowing of the Ni $3d$ band \cite{Braun-PRL-06}. While the above mentioned many-body
theories give an improved description of the electronic structure, the central 
question is, whether they also lead 
to a \emph{quantitative} 
agreement with experiments.

Angle-resolved photoemission spectroscopy (ARPES) is a powerful method to 
determine the spectral function and by 
comparison with the bare-particle band structure (usually approximated by DFT band structure calculations)
to obtain the self-energy \cite{Hufner-B-1995}. Moreover, the spin-resolved 
version of this method is very useful to disentangle the complex electronic structure of ferromagnets, in 
particular for systems with a strong overlap between majority and minority bands \cite{Jones-RMP-89}.
For ferromagnetic bcc Fe, which is the focus of the present work, several (spin-resolved)
ARPES studies have been presented in the literature 
\cite{Turner-PRB-84, Kisker-PRB-85, Santoni-PRB-91, Schafer-PRL-2004, Schafer-PRB-2005}.
  
In this letter we present (spin-resolved) ARPES spectra of ferromagnetic bcc Fe(110). 
These experimental results are compared to
theoretical many-body calculations (DMFT and 3BS) as mentioned above, 
including a full calculation of the one-step 
photoemission process. Our main conclusion is that a quantitative agreement, in particular
concerning the line widths, is not observed. This clearly demonstrates 
the demand for further non-local many-body 
theories.

Experiments have been performed at room temperature in normal emission using a 
hemispherical SPECS Phoibos 150 electron 
energy analyzer and undulator radiation from the
UE112-PGM1 beamline at BESSY II. The linear polarization of the photon beam 
could be switched from horizontal to vertical, which means that a more p- ({\bf A}$||$[1$\overline{1}$0]) 
or s- ({\bf A}$||$[001]) character of the incident light was achieved, respectively.
For spin analysis, a Rice University Mott-type spin polarimeter 
has been used, operated at 26 kV \cite{Burnett-RSI-94}. The angular resolution of the equipment was 
1\textordmasculineº, and the energy resolution was 100 meV. The Fe(110) 
surface has been prepared on W(110) by deposition of 20 ML Fe and post-annealing. The structural
quality of the films has been checked by low-energy electron diffraction. The Fe(110) film 
was remanently magnetized in the film plane along the [1$\overline{1}$0] easy axis. 

Theoretical calculations of the ARPES spectra have been performed within the framework of the 
Korringa-Kohn-Rostoker 
multiple scattering theory~\cite{Minar-PRB-05}. Spin-orbit coupling and exchange splitting
were treated on equal footing in a fully relativistic theory. 
To account for electronic correlations 
beyond the LSDA approximation, a 
site-diagonal, local  and complex energy-dependent self-energy $\Sigma_{DMFT}$ 
was introduced self-consistently 
\cite{Minar-PRL-05}. 
In most of the calculations presented for the comparison with the ARPES results, unless specified, 
we use for the averaged on-site Coulomb interaction $U$ a 
value $U$=1.5 eV which is within the experimental
value $U\approx$1 eV \cite{Steiner-92} and a value $U\approx$2 eV derived from theoretical 
studies \cite{Cococcioni-PRB-05,Chadov-EPL-08}.
It is usually accepted that the 
averaged on-site exchange interaction $J$ coincides with its atomic value $J\sim$0.9 eV \cite{Anisimov-PRB-91}.
Using this LSDA+DMFT approach the spectral function $A({E,\bf k})$ could be calculated for 
a semi-infinite lattice structure.
For a quantitative comparison with the measured ARPES spectra, it is inevitable to take into account
the wave-vector and energy dependent transition matrix elements 
calculated within the one-step model of photoemission 
(1SM) \cite{Braun-RPP-96}. This describes the excitation process, the transport of the 
photoelectrons to the surface as well as
the escape into the vacuum. For a quantitative description of surface states and resonances
a realistic approach for the surface barrier is essential.  
Finally, analogous calculations of the ARPES signal have been performed on the basis of the 3BS self-energy function.
The two calculations (3BS and DMFT) do not show big differences. We emphasize that 
in all the calculations the broadening
of the peaks due to final state effects is included.
 
Figures 1(a) and (b) display a comparison between spin-integrated ARPES data and 
theoretical LSDA+DMFT+1SM calculations of Fe(110) along the $\Gamma$N direction of the bulk Brillouin zone (BZ)
with p-polarization. The $k$ values were calculated from the
used photon energies ranging from 25eV to 100 eV, using an inner potential $V$=14.5 eV. 
Figure~2 displays analogous
spin-resolved ARPES data for p and 
s-polarized photons together with LSDA+DMFT+1SM calculations.

Near the $\Gamma$ point (k$\sim$0.06 $\Gamma$N), the intense peak close to the Fermi level corresponds to a 
$\Sigma_{1,3}^{\downarrow}$ minority surface resonance. Experimentally, its $\Sigma_{3}^{\downarrow}$ bulk 
component crosses the Fermi 
level at $k\sim$0.33 $\Gamma$N, leading to a reversal of the measured spin-polarization and to a strong 
reduction of the intensity at $k=$0.68 $\Gamma$N in the minority channel, in agreement with the 
theoretical results [Fig.~2(b) and (d)]. The peak at the binding energy
BE$\sim$0.7 eV, visible mainly for p-polarization 
in a large range of wave vectors between $\Gamma$ and N can
be assigned to almost degenerate $\Sigma_{1,4}^{\uparrow}$ bulk-like majority states [Figs.~1 and 2(a) and (c)].
\begin{figure}[tb]
\centering
\includegraphics [width=0.4\textwidth]{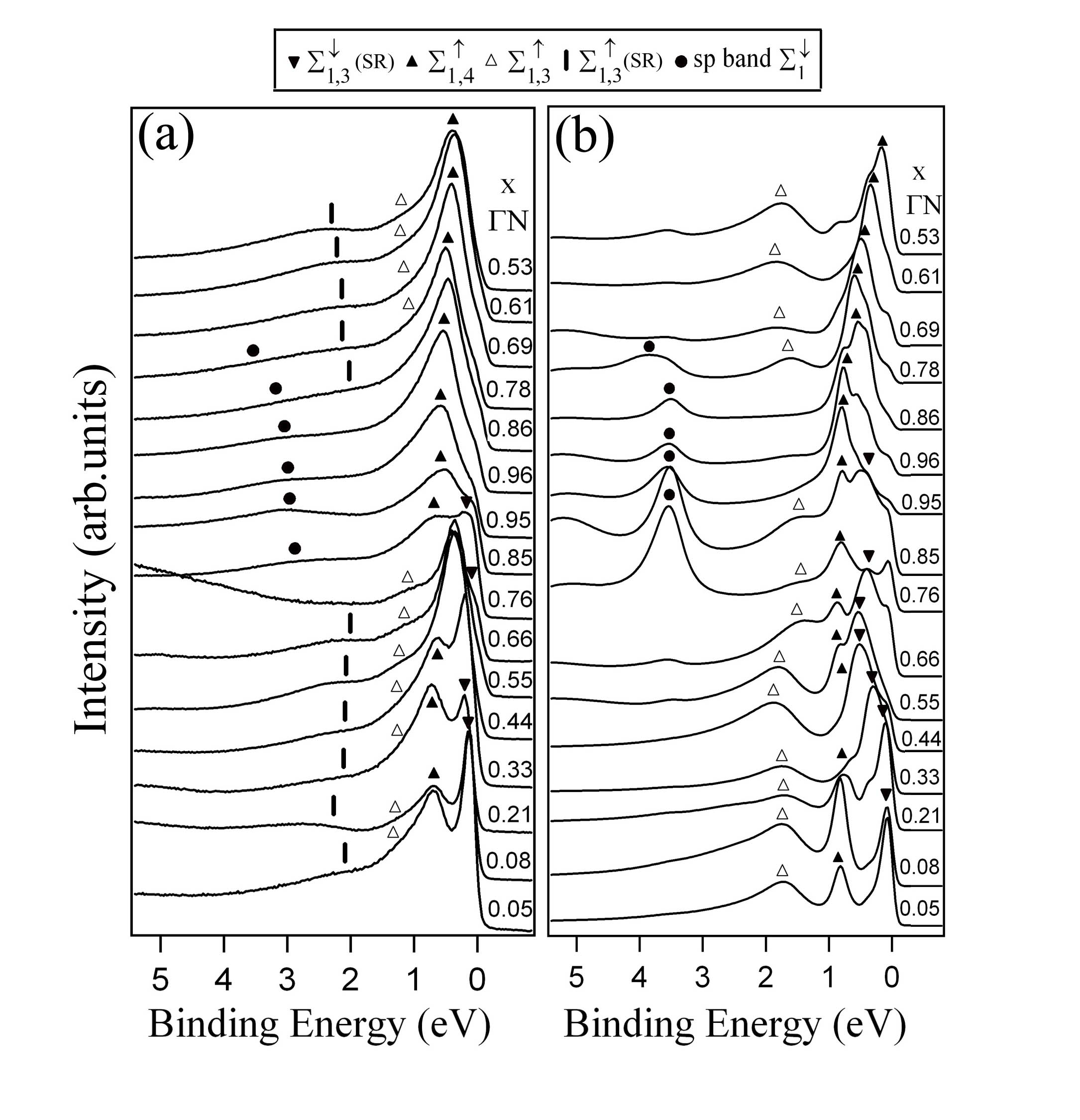}
\caption{(a) Experimental spin-integrated photoemission spectra of the Fe(110) surface 
measured with p-polarization in normal emission along the $\Gamma$N direction of the bulk Brillouin zone. 
The curves are labeled by the wave vectors in units of $\Gamma$N=1.55 $\AA^{-1}$. 
(b) Corresponding calculations obtained by the LSDA+DMFT+1SM method which 
includes correlations, matrix elements and surface effects.}
\label{fig:Fig1}
\end{figure}
For s-polarization [Fig.~2(b) and (d)], a $\Sigma_{3}^{\uparrow}$ feature at BE$\sim$1.1 eV  
dominates the spectrum at the $\Gamma$ point. For p-polarization its degenerate 
$\Sigma_{1}^{\uparrow}$ states form a shoulder around the same BE. 
The broad feature around 2.2 eV, visible at various $k$ points, 
but not at the N point, is related to a majority
$\Sigma_{1,3}^{\uparrow}$ surface state (see below). 
Around the N-point (0.76$\leq k\leq $1.0) and at BE$\geq$3 eV [Figs.~1(a) and 2(a)] we observe 
a $\Sigma_{1}^{\downarrow}$ band having strong $sp$ character. 
The pronounced difference between its theoretical 
and experimental intensity distributions can be attributed to the fact that 
in the present calculations only local Coulomb repulsion between d electrons is considered, without additional
lifetime effects for the sp bands. 
Finally, we notice that the background 
intensity of the spectrum at 
$k$=0.66 $\Gamma$N, corresponding to a photon energy of 55 eV, is strongly increasing by the appearance of the 
Fe 3p resonance.

\begin{figure}[tb]
\centering
\includegraphics [width=0.31\textwidth]{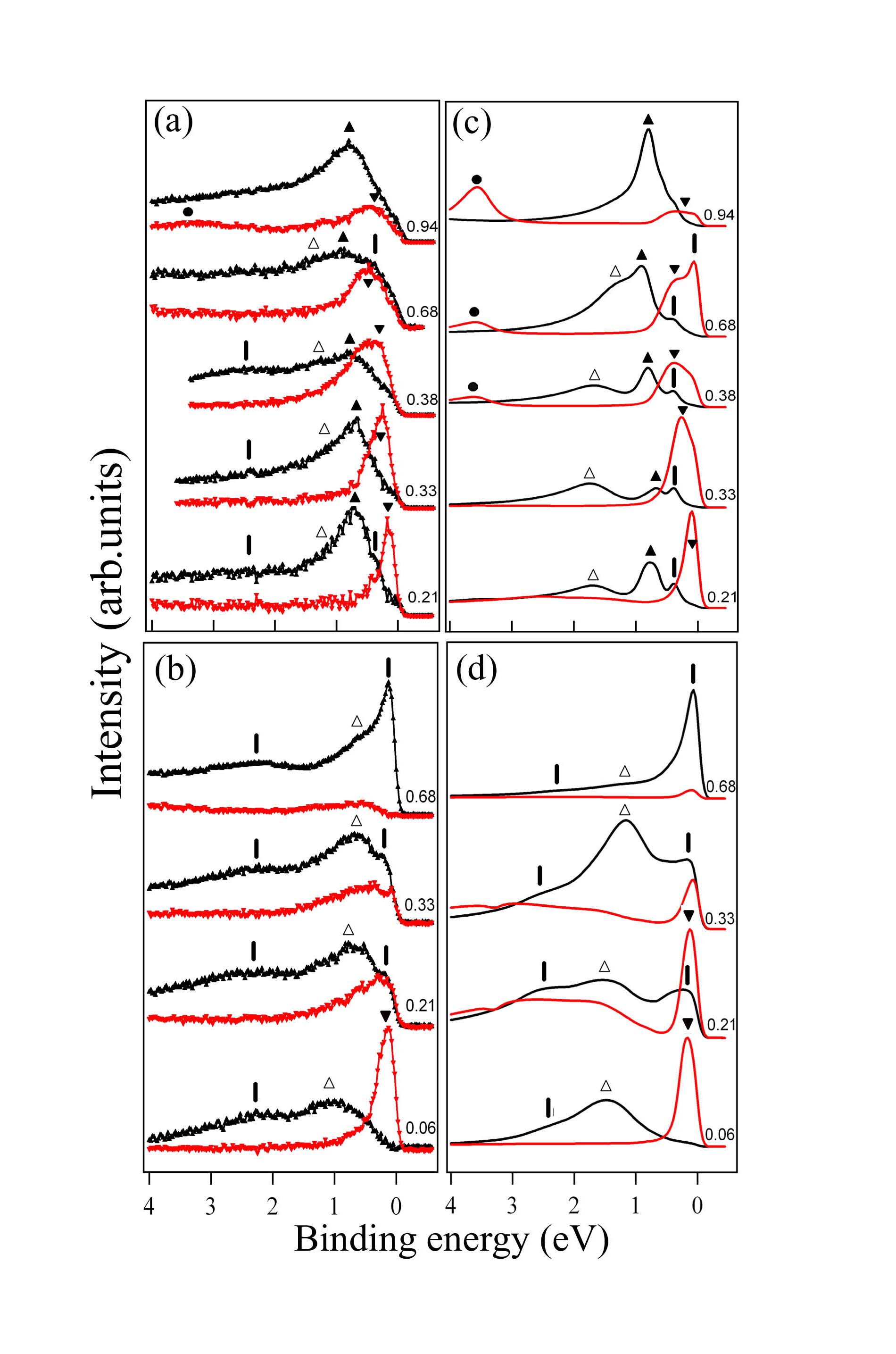}
\caption{(color online). Analogous data as in Fig.~1(a) and (b) but now spin-resolved.  
(a), (b): Experiment [black (red) upwards (downwards) triangles for majority (minority) electronic
states]. (c), (d): LSDA+DMFT+1SM theory [black (dark) and red (light) lines for majority and minority
electrons, respectively]. (a),(c) for p- and (b), (d) for s- polarization. The symbols
used for peak assignment are the same as in Fig.~1.} 
\label{fig:Fig2}
\end{figure}
\begin{figure}[tb]
\centering
\includegraphics [totalheight=0.42\textheight,width=0.35\textwidth]{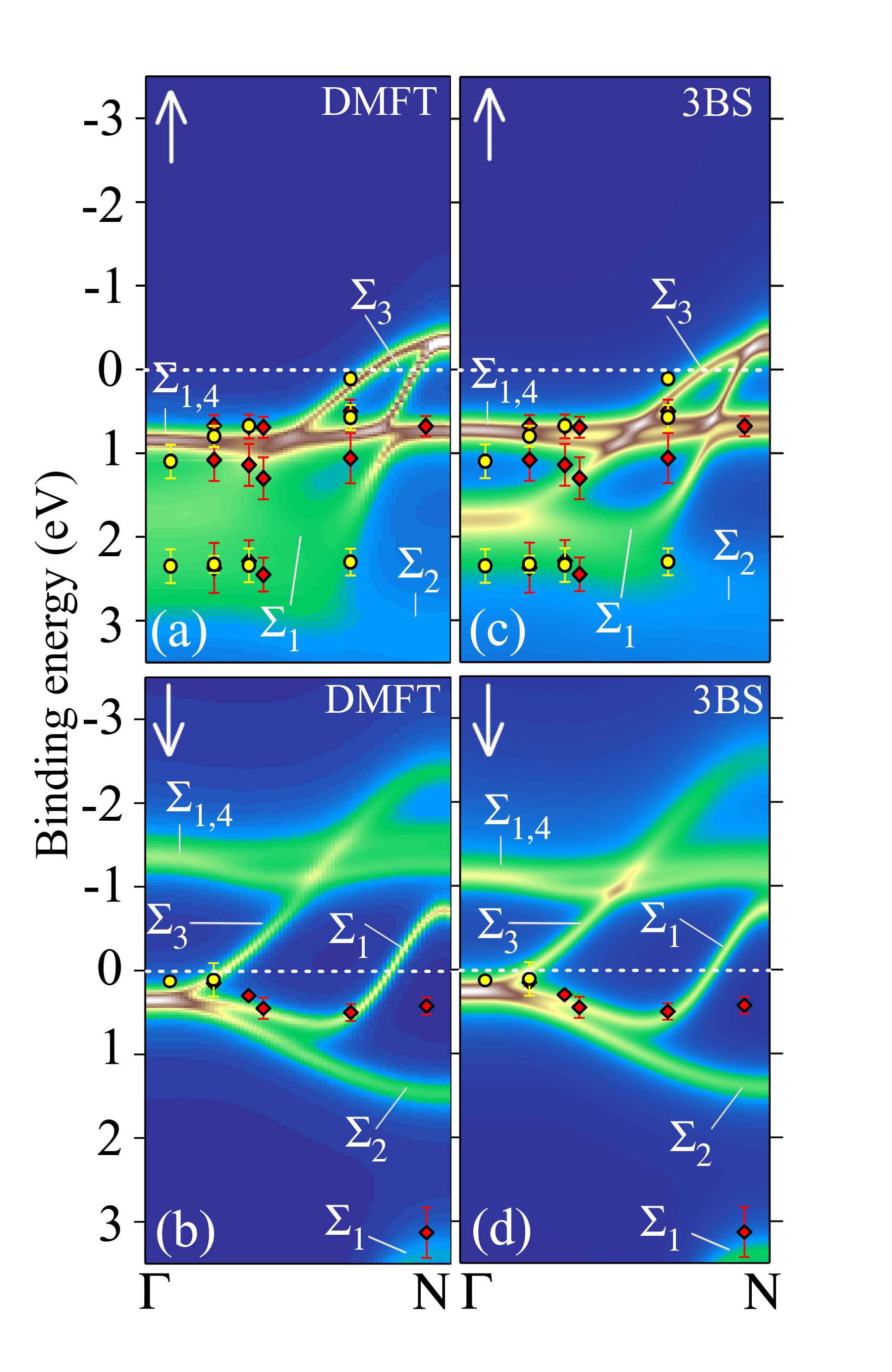}
\caption{(color online). Spectral functions of Fe(110) and photoemission peak positions obtained from the spin-resolved measurements for different polarizations ($\Diamond$ for horizontal and $\ocircle$ for vertical polarization). Results obtained by LSDA+DMFT [(a),(b)] and by LSDA+3BS [(c),(d)] methods for majority and minority electronic states,
respectively.}
\label{fig:Fig3}
\end{figure}
\begin{figure}[tbp]
\centering
\includegraphics [width=0.3\textwidth]{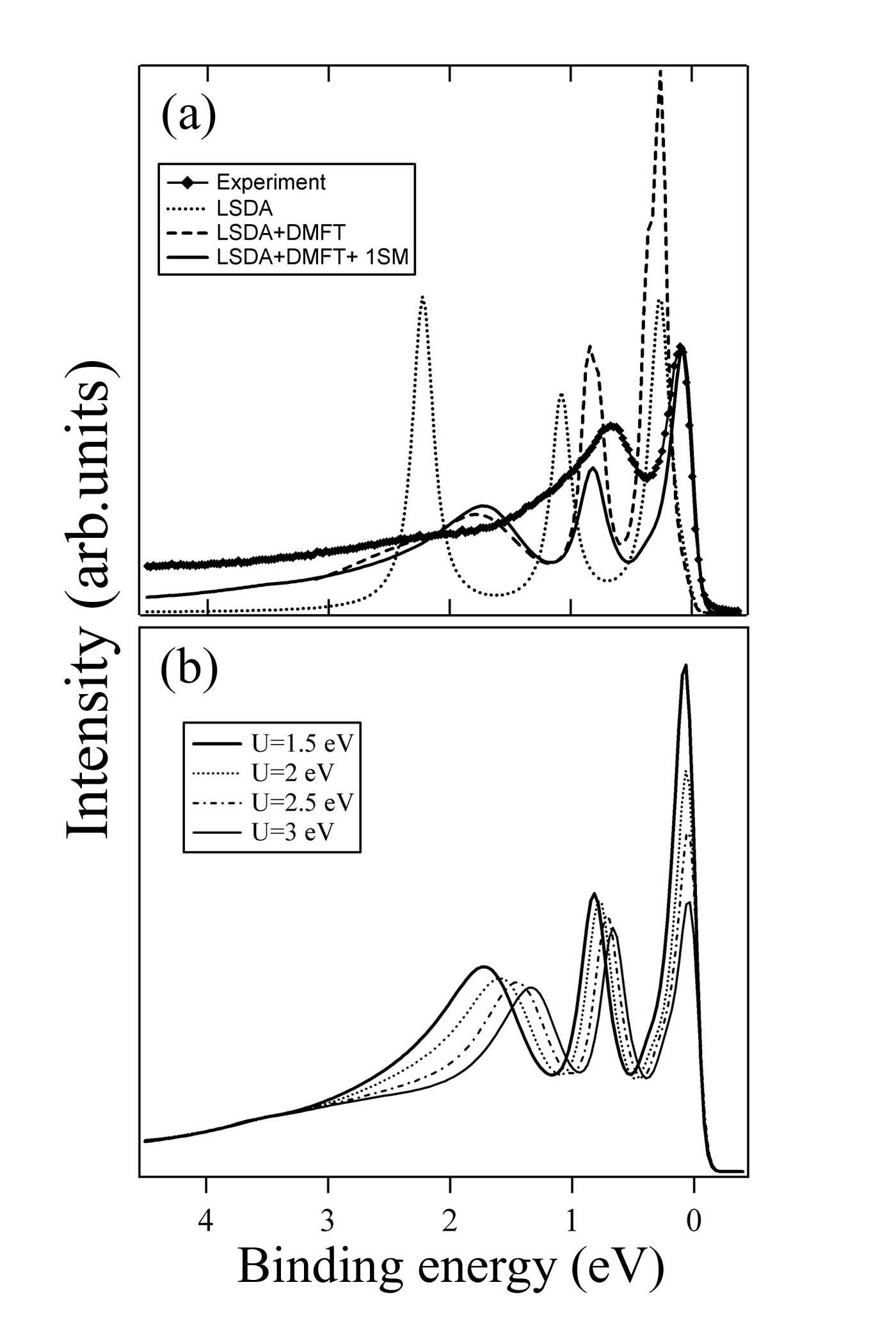}
\caption{(a) Comparison between the spin-integrated experimental spectra  
at the $\Gamma$ point for p-polarization with the single-particle LSDA-based calculation, 
the LSDA+DMFT spectra, and the LSDA+DMFT+1SM spectra.(b) LSDA+DMFT+1SM calculations for 
U=1.5 to 3 eV.}
\label{fig:Fig4}
\end{figure}

Comparing the experimental results from spin-integrated and spin-resolved ARPES measurements with 
LSDA+DMFT+1SM results, we obtain at low BE 
good agreement for many of the peak positions. 
This is also demonstrated in Fig.~3(a) and (b) were we compare the experimental peak positions 
with the LSDA+DMFT spectral function. Similar 
calculations based on the
LSDA+3BS scheme are compared with the experimental data in Fig.~3(c) and (d). Since the theoretical 
calculations do not show big differences, also the LSDA+3BS spectral function agrees well at low BE
with the experimental peak positions.

On the other hand, \emph{quantitative} agreement cannot be achieved for higher BE.
In particular, the calculated spectral weight near $\Gamma$ for the  $\Sigma_{1,3}^{\uparrow}$ bands
is in between the experimental features at 1.2 eV and 2.2 eV. Assuming negligible correlation
effects would move the calculated feature to the LDA value at BE=2.2 eV. Thus the experimental
peak at 2.2 eV could be assigned to the bulk $\Sigma_{1,3}^{\uparrow}$ bands. However, a complete neglect
of correlation effects in Fe would make the overall comparison between theory and experiment much worse. 
Thus we interpret the experimental peak at
BE=2.2 eV by a $\Sigma_{1,3}^{\uparrow}$ surface state in agreement with previous 
experimental and theoretical studies \cite{Kim-SSC-01}. Our theoretical results confirm this view since 
we clearly observe how changes in the surface barrier potential induce 
additional shifts in its BE position. Thus from the data shown in Fig.~3 we  
conclude that correlation effects
in the present calculations using $U$=1.5 eV are underestimated and that a stronger band narrowing
is needed to achieve agreement between theory and experiment.

Figure~4 demonstrates this for  
data close to the $\Gamma$ point ($k$=0.05 $\Gamma$N). Figure~4(a) compares the 
experimental spin-integrated ARPES spectra with LSDA calculations broadened with the experimental 
energy resolution, a LSDA+DMFT calculation, and a LSDA+DMFT+1SM calculation.
At low BE, perfect agreement between theory and experiment is achieved for the minority
$\Sigma_{1,3}^{\downarrow}$ surface resonance. For the bulk $\Sigma_{1,4}^{\uparrow}$ peak 
at BE=0.65 eV, the agreement 
is less satisfactory. This holds also for the $\Sigma_{1,3}^{\uparrow}$ peak which 
appears in LSDA at 2.2 eV. Due to correlation effects it is shifted in the experiment to BE$\approx$ 1.2 eV
causing in Fig.~4(a) at that 
energy a shoulder and in Fig.~2(b)
for $k=$0.06 $\Gamma$N a peak. The difference
between the ARPES data and LSDA calculations can be explained with a linear $\Re\Sigma$=0.7$E$ corresponding to
a mass enhancement $m^*/m_0$=1.7, where $m_0$ is the bare particle mass. 
This experimental mass renormalization in the energy range BE$\leq$ 1.8 eV should be contrasted 
to $m*/m_0\approx$1.25 derived for $U$=1.5 eV in the present work. 
One may speculate that the difference between the ARPES and the LSDA+DMFT+1SM peak positions could be reduced
by choosing a higher $U$ (e.g. $U\approx$ 4 eV).
This value, however, is outside the previously mentioned range and in addition 
we still are left with
the problem that the calculated width is far too small compared with the experimental value. 
It is remarkable that the width does not increase with increasing $U$ [Fig.~4(b)]. 
This can be explained in terms of
an energy dependent $\Im\Sigma$ and by the fact that with increasing $U$ the 
peak position moves to lower BE leading
to an almost constant $\Im\Sigma$. The energy dependence of $\Im\Sigma$ also 
leads to slightly asymmetric peaks in the
spectral function (Fig.~4(b)).

We have observed this additional broadening not only at the $\Gamma$ point but also at other $k$ values.
We have fitted the spin-resolved energy distribution curves also at various $k$ 
values by a sum of Lorentzians plus a
background. Although we are aware that such an evaluation is problematic 
due to the formation of asymmetric Lorentzians
discussed above, we have tried to extract $\Im\Sigma$ as a function of 
BE and $k$ (not shown). From this evaluation we obtain in a first approximation a $k$-independent 
$\Im\Sigma$, which is roughly a factor two bigger than the calculated $\Im\Sigma_{DMFT}$.

We point out that this additional broadening cannot be caused by final state effects \cite{Smith-PRB-93}, since 
those are fully taken into account in the 1SM calculations.
Furthermore, since the $3d$ bands show a rather flat dispersion, in particular close 
to the high symmetry points 
points of the BZ, the final state broadening can almost be neglected when compared with the 
experimental resolution. 
Moreover, the final state effects would
cause a broadening which for a small initial state dispersion, would be 
constant as a function of the BE. In this context, we mention that the 
broadening due to defect scattering is believed to be energy independent, as well. 

In summary, the comparison of (spin-resolved) ARPES data of ferromagnetic bcc Fe(110) to state of the art 
many-body LSDA+DMFT+1SM and LSDA+3BS+1SM calculations
indicates that these theories lead to an important improvement compared to LSDA. 
However, we have demonstrated that these theories, at least in the present implementation, 
underestimate the mass renormalization and in particular 
the scattering rates in Fe, a typical $3d$ multiband transition metal. This result demands for more 
refined many-body calculations, possibly with non-local schemes, which take into account non-local fluctuations
in multiband systems. These conclusions are also important for other correlated multiband systems such as 
Fe-pnictide high-$T_c$ superconductors.

\bibliographystyle{phaip}

\begin{thebibliography}{1}
\bibitem{Herring-M-66}
see e.g. C. Herring, in {\em Magnetism}, edited by G.T. Rado and H. Suhl, Vol. IV (Academic Press, New York, 1966).
\bibitem{Moruzzi-C-78}
V. L. Moruzzi {\it et al.}, {\em Calculated Electronic Properties of Metals} (Pergamon Press, 
Oxford, 1977).
\bibitem{Georges-RMP-96}
For a review, see A. Georges {\it et al.}, Rev. Mod. Phys. {\bf 68}, 13 (1996)
\bibitem{Calandra-PRB-94}
C. Calandra and F. Manghi, Phys. Rev. B {\bf 50}, 2061 (1994)
\bibitem{Monastra-PRL-02}
S. Monastra {\it et al.}, Phys. Rev. 
Lett. {\bf 23}, 236402 (2002)
\bibitem{Braun-PRL-06}
J. Braun {\it et al.}, Phys. Rev. Lett. {\bf 97}, 227601 (2006)  
\bibitem{Hufner-B-1995}
S. H\"ufner, {\em Photoelectron Spectroscopy} (Springer Verlag, Berlin, 1995)
\bibitem{Jones-RMP-89}
R. O. Jones and O. Gunnarsson, Rev. Mod. Phys. {\bf 61}, 689 (1989)
\bibitem{Turner-PRB-84}
A. M. Turner {\it et al.}, Phys. Rev. B {\bf 29}, 2986 (1984)
\bibitem{Kisker-PRB-85}
E. Kisker {\it et al.}, Phys. Rev. B {\bf 31}, 329 (1985)
\bibitem{Santoni-PRB-91}
A. Santoni and F. J. Himpsel, Phys. Rev. B {\bf 43}, 1305 (1991)
\bibitem{Schafer-PRL-2004}
J. Sch\"{a}fer {\it et al.}, Phys. Rev. Lett. {\bf 92}, 097205 (2004)
\bibitem{Schafer-PRB-2005}
J. Sch\"{a}fer {\it et al.}, Phys. Rev. B {\bf 72}, 155115 (2005)
\bibitem{Burnett-RSI-94}
G.C Burnett {\it et al.}, Rev. Sci. Instrum. {\bf 65} ,1893 (1994)
\bibitem{Minar-PRB-05}
J. Min\'ar {\it et al.}, Phys. Rev. B {\bf 72}, 045125 (2005) 
\bibitem{Minar-PRL-05}
J. Min\'ar {\it et al.}, Phys. Rev. Lett. {\bf 95}, 166401 (2005)
\bibitem{Steiner-92}
M.M. Steiner {\it et al.}, Phys. Rev. B {\bf 45}, 13272 (1992)
\bibitem{Cococcioni-PRB-05}
M. Cococcioni and S. de Gironcoli, Phys. Rev. B {\bf 71}, 035105 (2005)
\bibitem{Chadov-EPL-08}
S. Chadov {\it et al.}, EPL {\bf 82}, 37001 (2008) 
\bibitem{Anisimov-PRB-91}
V. I. Anisimov and O. Gunnarson, Phys. Rev. B {\bf 43}, 7570 (1991)
\bibitem{Braun-RPP-96}
J. Braun, Rep. Prog. Phys. {\bf 59}, 1267 (1996)
\bibitem{Kim-SSC-01}
H. -J. Kim {\it et al.}, Surf. Sci. {\bf 478}, 193 (2001)
%\bibitem{Grechnev-PRB-2007} 
%A. Grechnev {\it et al.}, Phys. Rev. B {\bf 76}, 035107 (2007)
\bibitem{Smith-PRB-93}
N. V. Smith {\it et al.}, Phys. Rev. B {\bf 47}, 15476 (1993)
 \end{thebibliography}

\end{document}